\documentclass[prx,aps,twocolumn,superscriptaddress]{revtex4-1}

\usepackage[pdftex]{graphicx}
\usepackage{amsbsy,amssymb,amsmath,bm,mathtools}

\usepackage{xspace}
\usepackage{hyperref}

\usepackage{bm}
\usepackage{color}

\begin{document}


\title{Nonequilibrium dynamics of superconductivity in the attractive Hubbard model}

\author{Gia-Wei Chern}
\affiliation{Department of Physics, University of Virginia, Charlottesville, VA 22904, USA}

\author{Kipton Barros}
\affiliation{Theoretical Division and CNLS, Los Alamos National Laboratory, Los Alamos, New Mexico 87545, USA}

\date{\today}

\begin{abstract}
We present a framework of semiclassical superconductivity (SC) dynamics that properly includes effects of spatial fluctuations for the attractive Hubbard model.  We consider both coherent and adiabatic limits. To model the coherent SC dynamics, we develop a real-space von~Neumann equation based on the time-dependent Hartree-Fock-Bogoliubov theory. Applying our method to interaction quenches in the negative-$U$ Hubbard model, we show that the relaxation of SC order at weak coupling is dominated by Landau-damping. At strong coupling, we find a two-stage relaxation of the pairing field: a collapse of the synchronized oscillation of Cooper pairs due to spatial inhomogeneity, followed by a slow relaxation to a quasi-stationary state. SC dynamics in adiabatic limit is described by a quantum Landau-Lifshitz equation with Ginzburg-Landau relaxation. Numerical simulations of the pump-probe process show that long time recovery of the pairing field is dominated by defects dynamics. Our results demonstrate the important role of spatial fluctuations in both limits.
\end{abstract}

\maketitle

The nonequilibrium dynamics of superconductivity~(SC) subject to an external stimulation has been intensively studied for some time~\cite{kopnin01,volkov74,barankov04,andreev04,szymanska05,yuzbashyan06a,barankov06,yuzbashyan06}. This interest has recently been renewed by remarkable pump-probe experiments reporting the observation of the collective amplitude mode~\cite{matsunaga13,matsunaga14}. 
Prior work largely studies two physical limits determined by the relation between the quasiparticle relaxation time $\tau_{\epsilon}$ and the relaxation time $\tau_\Delta$ of SC order parameter. The SC dynamics in the collisionless limit, $\tau_\epsilon \gg \tau_\Delta$, can be described using a time-dependent self-consistent field approach. An interesting phenomenon in this regime is the collective Rabi oscillations of the  order parameter~\cite{barankov04,yuzbashyan06a}.  In the opposite, adiabatic limit, $\tau_{\epsilon} \ll \tau_{\Delta}$, the dynamics of the pairing field is usually described by time-dependent Ginzburg-Landau (TDGL) equation~\cite{binder73,hohenberg77}. TDGL has recently been employed to simulate the out-of-equilibrium dynamics of superconductors in pump-probe experiments~\cite{madan16,kennes17}. 

Numerical simulations based on TDGL also provide useful insights on the dynamical inhomogeneity of nonequilibrium superconductivity~\cite{karra98,kopnin99}, for example, the formation of topological defects by rapid thermal quenches as described in the Kibble-Zurek scenario~\cite{kibble76,zurek96}. However, contrary to numerous large-scale TDGL simulation studies, effects of spatial fluctuations in the collisionless limit are rarely addressed in most numerical studies, even though earlier calculations~\cite{dzero09,foster10} have demonstrated dramatic effects of spatial inhomogeneity in out-of-equilibrium superconductivity.

In this paper, we present a theoretical framework for SC dynamics that properly includes the effects of spatial fluctuations in both the collisionless and adiabatic limits. To model the collisionless (coherent) limit, we develop a real-space formulation of the time-dependent Hartree-Fock-Bogoliubov (TDHFB) theory, which is particularly suitable for the negative-$U$ Hubbard model. Numerical solution shows two distinct dynamical regimes for the evolution of the pairing field. In particular, we demonstrate an inhomogeneity-induced collapsing of the synchronized oscillation of the SC order parameter at large coupling. To model the adiabatic limit, we use a {\em real-space} Anderson pseudo-spin representation~\cite{anderson58} and show that the SC dynamics is described by a quantum Landau-Lifshitz equation with Ginzburg-Laudau relaxation. 

We consider the Hubbard model with an attractive on-site interaction on the square lattice~\cite{scalettar89,moreo91,micnas90,paiva10}:
\begin{eqnarray}
	\label{eq:H0}
	\mathcal{H} = -\!\sum_{ ij , \alpha}  t_{ij} \, c^{\dagger}_{i,\alpha} c^{\;}_{j,\alpha} + U \sum_{i}  n_{i, \uparrow} n_{i, \downarrow} 
	 - \sum_i  \mu n_i,
\end{eqnarray}
Here $c^\dagger_{i, \alpha}$ is the creation operator of fermions with spin $\alpha = \uparrow, \downarrow$ at site-$i$, $n_{i, \alpha} = c^\dagger_{i,\alpha} c^{\;}_{i, \alpha}$ is the fermion number operator, and $n_i = n_{i, \uparrow} + n_{i, \downarrow}$. We assume on-site attraction, $U < 0$. A nonzero chemical potential $\mu$ tunes the fermion density away from half-filling. The negative-$U$ Hubbard model provides a simple platform for investigating the crossover between the BCS and the Bose-Einstein condensation (BEC) regimes of fermionic superfluid. The model is known to have an enhanced O(3) symmetry at half-filling ($\mu = 0$), corresponding to the co-existence of SC and charge density wave (CDW) orders. Away from $\mu = 0$, this degeneracy is lifted and the SC state is energetically more favorable~\cite{scalettar89,moreo91}. 

To model the coherent SC dynamics, we develop a real-space equation of motion approach based on the TDHFB theory, which is equivalent to the time-dependent Bogoliubov-de~Gennes equations~\cite{degennes66}.  By introducing a pairing field $\Delta_i$ and on-site density $\rho_i$, we perform the Hubbard-Stratonovich transformation to obtain the following BdG Hamiltonian: 
\begin{eqnarray}
	\label{eq:H_BdG}
	& & \mathcal{H}_{\rm BdG} = -\!\sum_{ ij , \alpha}  t_{ij} c^{\dagger}_{i,\alpha} c^{\;}_{j,\alpha}  
	+ U \sum_i ( \Delta_i c^\dagger_{i, \uparrow} c^\dagger_{i, \downarrow} + \mbox{h.c.}) \nonumber \\
	& & \qquad \,\, + \sum_i (U \rho_i - \mu) \, n_i  + U \sum_i (|\Delta_i|^2 + \rho_i^2).
\end{eqnarray}
Self-consistency requires that $\Delta_i = \langle c_{i, \uparrow} c_{i, \downarrow} \rangle$ and $\rho_i = \langle c^\dagger_{i, \alpha} c^{\;}_{i, \alpha} \rangle$ for both spins $\alpha = \uparrow, \downarrow$, where the average $\langle \cdots \rangle$ is computed using the BdG Hamiltonian in Eq.~(\ref{eq:H_BdG}). This auxiliary field Hamiltonian {\em without the self-consistency constraint} is usually the starting point of determinant quantum Monte Carlo (DQMC) method, which can be applied to the negative-$U$ Hubbard model thanks to absence of the sign-problem~\cite{scalettar89,moreo91,paiva10}. Equation~(\ref{eq:H_BdG}) is also the basis of classical Monte Carlo method assuming {\em static} auxiliary fields~\cite{mayr05,conduit11,datta14,tarat15}. Although this method neglects quantum fluctuations in imaginary time, it does take into account thermal and spatial fluctuations of the order-parameter field, which are not included in a conventional mean-field treatment. Indeed, results obtained from the static auxiliary-field Monte Carlo agree remarkably well with that of DQMC simulations~\cite{datta14,tarat15}.

The dynamics of the pairing field in the TDHFB theory is given by the Heisenberg equation of motion, $d\Delta_i / dt = (i/\hbar) \langle [\mathcal{H}_{\rm BdG}, c_{i, \uparrow} c_{i, \downarrow}] \rangle$. In fact, the on-site density and the SC order are essentially the diagonal elements of the normal and anomalous density matrices, respectively. Namely, $\rho_{ij} \equiv \langle c^\dagger_{j, \uparrow} c^{\;}_{i, \uparrow} \rangle =  \langle c^\dagger_{j, \downarrow} c^{\;}_{i, \downarrow} \rangle$ and $\Delta_{ij} \equiv \langle c_{j,\uparrow} c_{i, \downarrow} \rangle$ for a non-magnetic superfluid, which is our primary interest.  A description of the SC dynamics requires the time evolution of both $\rho_{ij}$ and $\Delta_{ij}$, which is governed by the generalized von~Neumann equation, $d D/dt = (H_{\rm BdG} D - D H_{\rm BdG}^\dagger)/i \hbar$, where $D$ is a generalized single-particle density matrix that includes normal as well as anomalous components.  Explicitly,  
\begin{eqnarray}
	\label{eq:drho_dt}
	i\hbar \frac{d\rho_{ij}}{dt} &=& \sum_k (t_{ik} \rho_{kj} - \rho_{ik} t_{kj} ) + U (\rho_i - \rho_j) \rho_{ij} \nonumber \\
	& & \quad +U \left(\Delta_i \Delta^*_{ij} -  \Delta_{ij} \Delta^*_j \right), \\
	\label{eq:dDelta_dt}
	i\hbar \frac{d \Delta_{ij}}{dt} & = & \sum_k (t_{ik} \Delta_{kj} + \Delta_{ik} t_{kj}) + [U ( \rho_i + \rho_j) - 2\mu] \Delta_{ij} \nonumber \\
	& & \quad + U \delta_{ij} \Delta_i - U (\rho_{ij} \Delta_j + \rho_{ji} \Delta_i).
\end{eqnarray}
Similar equations have been used to model the superfluid dynamics of cold atoms~\cite{bulgac17,boudjemaa11} using time-dependent density-functional theory~\cite{bulgac13}.

The above TDHFB can be further simplified if we ignore the spatial fluctuations of SC and CDW order parameters; we shall refer to this approximation as the time-dependent mean-field (TDMF) method.  The assumption of translation invariance allows for direct solution of Eqs.~(\ref{eq:drho_dt}) and (\ref{eq:dDelta_dt}) in the Fourier representation. Furthermore, the number of dynamical variables reduces from $\mathcal{O}(N^2)$ to $\mathcal{O}(N)$, where $N$ is the number of lattice sites. This TDMF method becomes exact in the special case of a BCS superconductor~\cite{barankov04,yuzbashyan06a,barankov06,yuzbashyan06}, and has been widely employed to study quenched superconductors~\cite{foster14,peronaci15,sentef17}, including the negative-$U$ Hubbard model~\cite{bunemann17,mazza17}.

\begin{figure}[t]
\includegraphics[width=0.99\columnwidth]{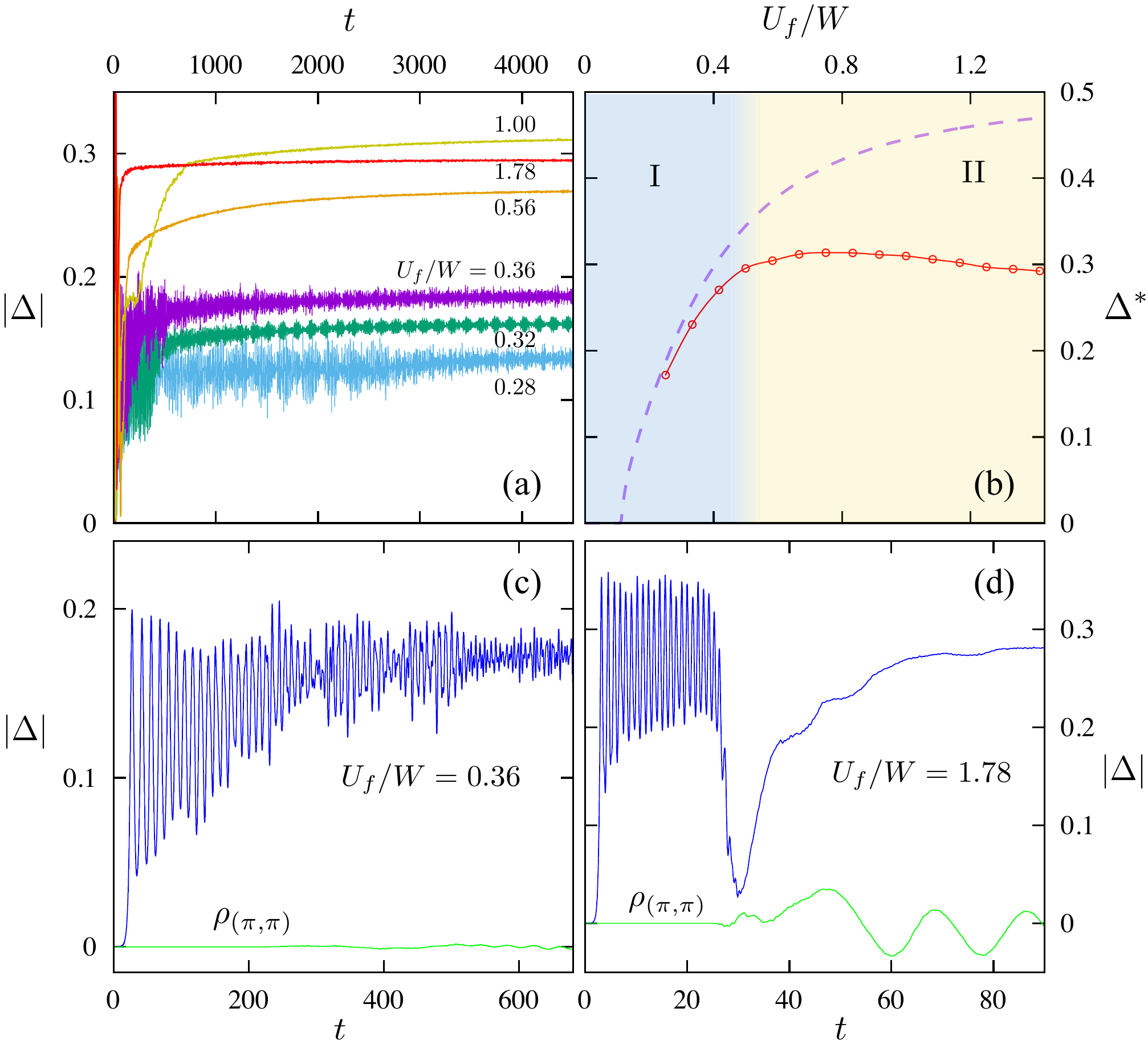}
\caption{(Color online)  
\label{fig:traces} (a) The time dependence of  the spatially averaged SC order parameter $\overline\Delta$ for varying final $U_f$. The initial strength of the on-site attraction is $|U| = U_i = 1.2 t_{\rm nn}$ and $W = 8 t_{\rm nn}$ is the bandwidth. (b) The quasi-stationary value of $\Delta^*$ vs $U_f$, the dashed line shows the equilibrium SC order when $|U| = U_f$. Panels (c) and (d) shows the SC and CDW order versus time at short time scales for $U_f/W = 0.36$ and 1.78, respectively.
}
\end{figure}

Here we apply the real-space TDHFB formulation to simulate quenches of the interaction strength in the negative-$U$ Hubbard model at zero temperature. We consider a time-dependent on-site attraction, $U(t) = -U_i$ for $t < 0$, which suddenly increases in magnitude, $U(t) = -U_f$ for $t \ge 0$. A chemical potential $\mu = 0.05 \, t_{\rm nn}$ is used to stabilize a small SC order in the initial state. We have also included a tiny random on-site potential of order $\epsilon_i \sim 10^{-7} t_{\rm nn}$. This small initial perturbation is introduced to examine whether the out-of-equilibrium states are stable against inhomogeneities. The time dependence of the SC and CDW order parameters shown in Fig.~\ref{fig:traces}(a) exhibits two distinct dynamical behaviors depending on the strength of the final $U_f$. The quasi-stationary value $\Delta^*$ at large $t$ is plotted in Fig.~\ref{fig:traces}(b) as a function of $U_f$; also shown for comparison is the equilibrium $\Delta$ at the corresponding $U = -U_f$. The crossover from the weak to strong coupling regimes roughly corresponds to the maximum of~$\Delta^*$. We discuss the characteristic features of the two dynamical regimes below.

For small $U_f$, such as the case shown in Fig.~\ref{fig:traces}(c), the pairing order parameter $\overline{\Delta}$ oscillates with a frequency that is proportional to $U_f$ and exhibits collisionless dephasing. The dephased oscillation mainly results from the energy exchange between the collective mode, i.e. the SC order parameter $\overline{\Delta} = \sum_{\mathbf k} \Delta_{\mathbf k} / N$, and the individual Cooper pairs $\Delta_{\mathbf k} = \langle \tilde c_{\mathbf k, \uparrow} \tilde c_{-\mathbf k, \downarrow} \rangle$ in {\em momentum} space, as described in the Landau-damping mechanism. Moreover, we find that the pairing field $\Delta(\mathbf r)$ in this weak-coupling regime shows weak inhomogeneity, which is dominated by long-wavelength fluctuations.  The scenario described here is similar to that of interaction quenches of BCS superconductors~\cite{barankov04,yuzbashyan06a,barankov06,yuzbashyan06}. This is also consistent with the fact that the SC order in the weak-coupling regime of the negative-$U$ Hubbard model is better described by a BCS-type model with a large coherence length. At longer time scales, the CDW order $\rho_{(\pi, \pi)}$ emerges as the pairing order parameter settles to its quasi-stationary value.

\begin{figure}[t]
\includegraphics[width=0.95\columnwidth]{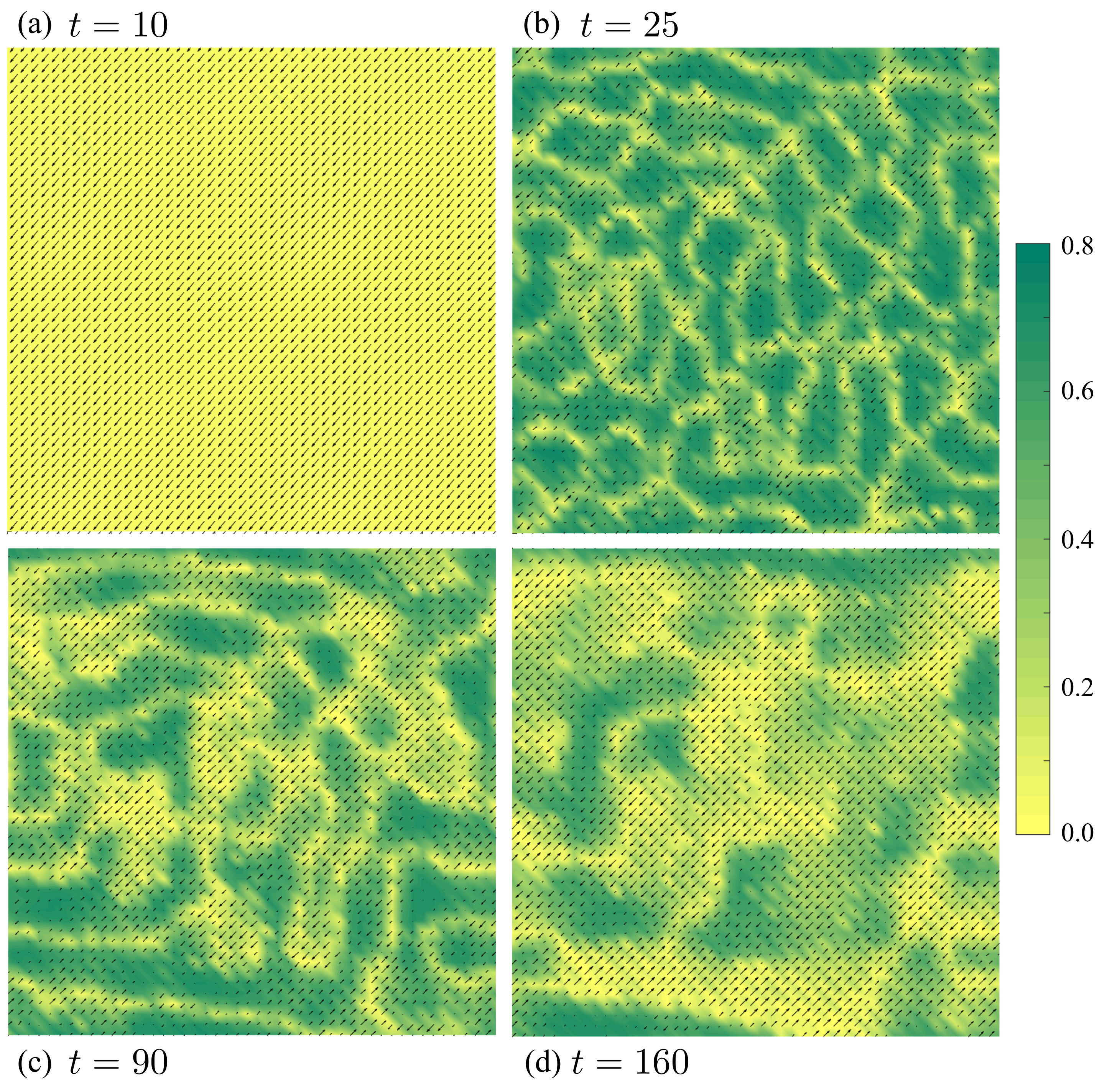}
\caption{(Color online)  
\label{fig:snapshots1} Snapshots of pairing field for quench from $U_i = 0.15W$ to $U_f = 1.78 W$. The arrow denotes the complex SC order parameter, i.e. $({\rm Re}\Delta_i, {\rm Im}\Delta_i)$, while the gradient color indicates the absolute value of on-site charge density $|\rho_{i, i}|$. The simulations are done on a $48\times 48$ lattice.
}
\end{figure}

As $U_f$ increases,  the pairing order parameter initially exhibits an oscillation of relatively constant amplitude as demonstrated in Fig.~\ref{fig:traces}(d). The spatial distribution of $\Delta_i$ is rather uniform during this initial stage; see Fig.~\ref{fig:snapshots1}(a). This coherent oscillation in the short timescales resembles the phase-locked regime discussed in the quantum quench of BCS superconductors~\cite{barankov04,yuzbashyan06a,barankov06,yuzbashyan06}. Importantly, we find that the synchronized oscillation of Cooper pairs is unstable against disorder: any small initial 
inhomogeneity is amplified after the interaction quench, giving rise to collapse of the synchronized oscillation as shown in Fig.~\ref{fig:traces}(d). Furthermore, the collapse of the initial coherent oscillation is accompanied by the emergence of highly inhomogeneous SC and CDW fields; see Figs.~\ref{fig:snapshots1}(b)--(d).

The emergence of spatially inhomogeneous SC field most likely results from the parametric instability as first pointed out by Dzero {\em et al.} in Ref.~\cite{dzero09}. It is worth noting that parametric instability and similar turbulence phenomena have been discussed in the dynamics of other nonlinear systems~\cite{suhl57,zakharov75,lvov73,bunkov06}. In this so-called Cooper pair turbulence scenario~\cite{dzero09}, spatial modulations develop through parametric excitations of pairing modes $\Delta_{\pm\mathbf k}$ with opposite momenta. The initial exponential growth of these modes is suppressed by higher-order scattering processes, giving rise to a  state which can be viewed as a random superposition of wave packets of the pairing order parameter. Importantly, this instability takes place when the SC coherence length is smaller than the system size~\cite{dzero09}. This indeed corresponds to the strong coupling case of negative-$U$ Hubbard model, where the SC state is better described by a BEC of preformed fermion pairs $\Delta_i = \langle c_{i, \uparrow} c_{i, \downarrow} \rangle$ of relatively small coherence length. Finally, we note that while the overall picture seems consistent with the Cooper-pair turbulence picture, further numerical investigations, such as the momentum distribution of the pairing modes, are required in order to fully characterize this scenario.

Now we turn to the {\em adiabatic} limit, $\tau_{\epsilon} \ll \tau_{\Delta}$, of the TDHFB and show that the pairing order parameter follows a novel Ginzburg-Landau-Lifshitz dynamics. We first employ the well known canonical transformation $c_{i, \uparrow} \to \tilde c_{i, \uparrow}$ and $c_{i, \downarrow} \to e^{i \mathbf Q \cdot \mathbf r_i} \,\tilde c_{i, \downarrow}^\dagger$, to map Hamiltonian~(\ref{eq:H0}) to a positive-$U$ Hubbard model at half-filling in a uniform magnetic field $H_z = 2 \mu$~\cite{scalettar89,moreo91}. Here $\mathbf Q = (\pi, \pi)$ and the phase factor $e^{i \mathbf Q\cdot \mathbf r_i} = \pm 1$ describes a checkerboard pattern on the square lattice. In the transformed representation, the order parameters can be conveniently grouped into a pseudo-spin $\mathbf T_i$ with components:
\begin{eqnarray}
	T_i^x + i T_i^y =  \Delta_i\, e^{i \mathbf Q \cdot \mathbf r_i} , \qquad  T_i^z = (\rho_i - 1)/2.
\end{eqnarray}
This essentially maps the combined SC/CDW order into a spin-density wave (SDW) order in a repulsive Hubbard model. The vector $\mathbf T_i$ is the real-space version of the pseudo-spin introduced by Anderson for the dynamics of BCS superconductors~\cite{anderson58}. In the large $U$ limit, the BdG Hamiltonian reduces to the Heisenberg exchange interaction, and the pseudo-spins  satisfy the LL dynamics in {\em real} space: $d\mathbf T_i / dt = J \sum_j \mathbf T_j \times \mathbf T_i - \mu \hat{\mathbf z} \times \mathbf T_i$~\cite{burkov08}.

Generalization of the semiclassical SDW dynamics to the intermediate $U$ regime has been recently developed in Ref.~\cite{chern17}. Here we briefly outline the formulation in our context. The time evolution of the pseudo-spin $\mathbf T_i$ is governed by the conservation of pseudo angular momentum:
\begin{eqnarray}
	\label{eq:dTdt}
	\frac{d\mathbf T_i}{dt} = -\frac{i}{2} \sum\nolimits_j t_{ij} \bm\sigma_{\beta\alpha} \left(\tilde{\rho}_{i\alpha, j\beta} - \tilde{\rho}_{j\alpha, i\beta} \right),
\end{eqnarray}
where  $\tilde{\rho}_{i\alpha, j\beta} \equiv \langle \tilde c^\dagger_{j\beta} \tilde c^{\;}_{i\alpha} \rangle$ are the reduced density matrix elements for the transformed $\tilde c$ fermions, and are related to $\rho_{ij}$ and $\Delta_{ij}$ in the original representation. 
The dynamics of the density matrix $\tilde\rho$ is again governed by von~Neumann equation $d\tilde \rho/dt = i [\tilde \rho, \tilde H_{\rm BdG} ]$, which is equivalent to Eqs.~(\ref{eq:drho_dt}) and~(\ref{eq:dDelta_dt}). Here $\tilde H_{\rm BdG}$ is the single-particle transformed Hamiltonian; symbolically, $\tilde{\mathcal{H}}_{\rm BdG} = \sum \tilde c^{\dagger}\, \tilde H_{\rm BdG} \, \tilde c^{\;}$.

\begin{figure}[t]
\includegraphics[width=0.99\columnwidth]{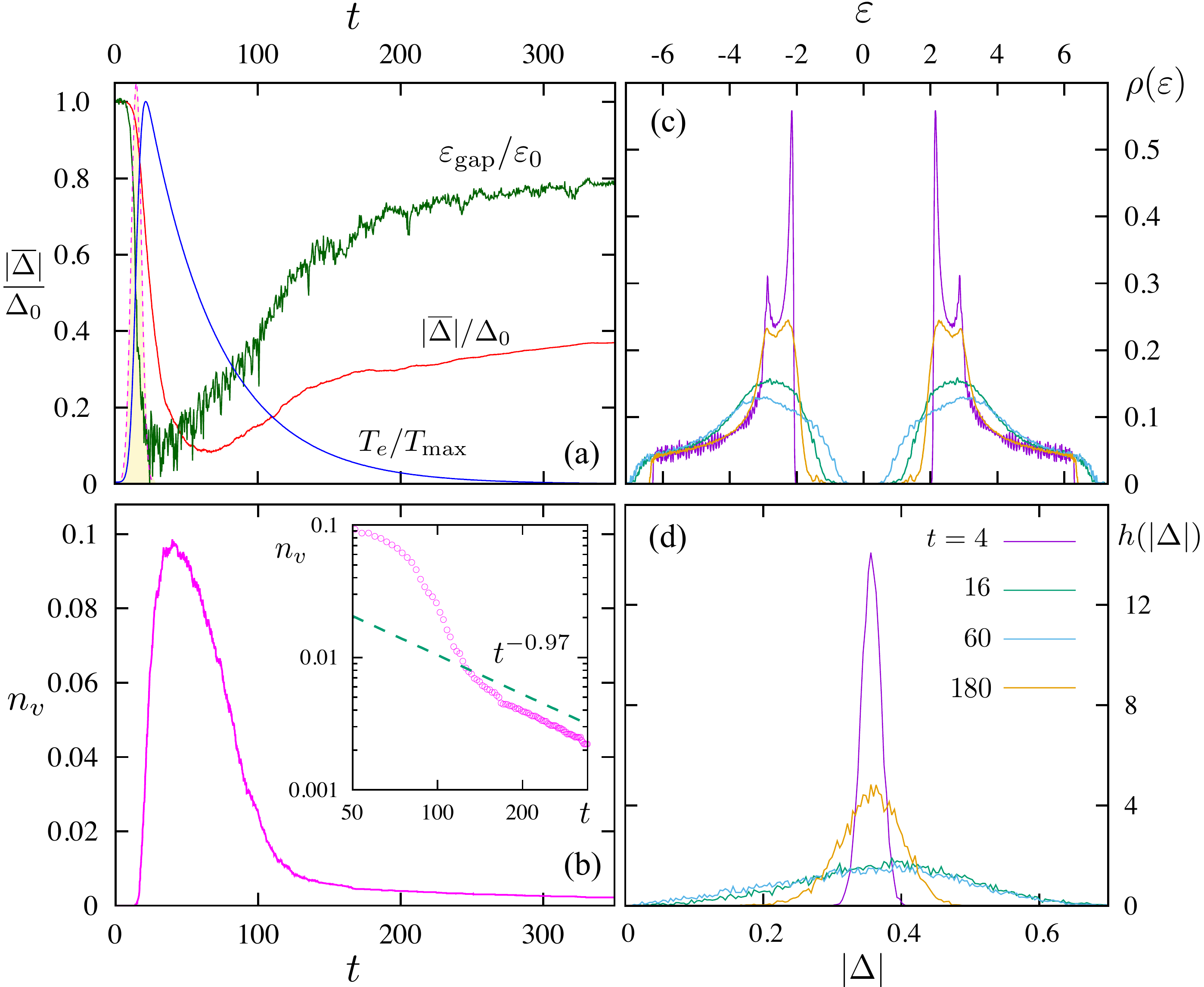}
\caption{(Color online)  
\label{fig:pump-probe} (a) Time dependence of the effective electron temperature $T_e$, electron energy gap $\varepsilon_{\rm gap}$, and SC order parameter $\overline \Delta$. The dashed line indicates the temporal profile of a pump pulse. $\varepsilon_0$ is the gap at $t = 0$. (b) Vortex density $n_v$ as a function of time; the inset shows the same dependence in log-log scale. (c) Electron density of states, and (d) probability distribution of pairing amplitude $|\Delta|$ at varying simulation times. The simulated lattice has $N = 120\times 120$ sites.}
\end{figure}

In the adiabatic limit, electrons are assumed to quickly relax to the equilibrium state of the instantaneous $\mathcal{H}_{\rm BdG}$. Consequently, we approximate the density matrix on the right-hand side of Eq.~(\ref{eq:dTdt}) by $\tilde\rho^{(0)}$ that describes the equilibrium electron liquid of the instantaneous Hamiltonian, i.e. $[\tilde \rho^{(0)}, \tilde{H}_{\rm BdG}] = 0$. Substituting $\tilde \rho^{(0)}$ in Eq.~(\ref{eq:dTdt}) leads to our quantum Landau-Lifshitz dynamics (QLLD) method~\cite{chern17}. We include a Ginzburg-Landau type damping and stochastic driving force to obtain
\begin{eqnarray}
	\label{eq:QLLD}
	\frac{d\mathbf T_i}{dt} = -\mathbf T_i \times \frac{\partial \langle \tilde{\mathcal{H}}_{\rm BdG} \rangle}{\partial \mathbf T_i} 
		- \gamma_{\mu} \frac{\partial \langle \tilde{\mathcal{H}}_{\rm BdG} \rangle}{\partial \mathbf  T^{\mu}_i} + \bm\xi_i(t),
\end{eqnarray}
where $\langle \tilde{\mathcal{H}}_{\rm BdG} \rangle = {\rm Tr}(\tilde{\rho}^{(0)} \tilde H_{\rm BdG})$,  $\mu = x, y, z$ denotes the components of pseudo-spin, $\gamma_x = \gamma_y$ and $\gamma_z$ are the damping constants of SC and CDW order parameters, respectively, and $\bm\xi_i(t)$ is a $\delta$-correlated fluctuating force satisfying $\langle \bm\xi_i(t) \rangle = 0$ and $\langle \xi^\mu_i(t) \xi^\nu_i(t') \rangle = 2 \gamma_\mu k_B T \delta_{ij} \delta_{\mu\nu} \delta(t - t')$. Note that QLLD requires solution of the equilibrium density matrix $\tilde{\rho}^{(0)}$ at {\em every} time-step, in analogy to Born-Oppenheimer quantum molecular dynamics~\cite{marx00}. Rather than direct diagonalization of $\tilde{\mathcal{H}}_{\rm BdG}$, we use the kernel polynomial method~\cite{silver94,weisse06} with gradient-based probing~\cite{barros13,wang17} to estimate $\tilde\rho^{(0)}$, and thus effective forces, at a cost that scales linearly with system size.

We next apply the QLLD to investigate the ultrafast relaxation of SC order subject to a short laser pulse. For simplicity, we assume that the effect of the pump pulse is to inject energy to the electrons, which quickly equilibrate to a state characterized by temperature $T_e$. This is consistent with our adiabatic approximation for the SC dynamics. The time dependence of the effective electron temperature is governed by the rate equation $C dT_e/dt = -G (T_e - T_L) + Q(t)$~\cite{anisimov74}, where $C$ is the heat-capacity of the electron liquid, $G$ is the coupling to the lattice, $T_L$ is the lattice temperature, and $Q(t) \propto \exp[-(t - t_p)^2 / w^2]$ is the heat source due to the pump pulse. We further assume that $T_L \approx 0$ throughout the relaxation process. The resultant $T_e(t)$, shown  in Fig.~\ref{fig:pump-probe}(a), then controls the magnitude of the stochastic noise $\bm\xi(t)$ in our QLLD simulations of Eq.~(\ref{eq:QLLD}). We use the parameters, $U = -3$, damping $\gamma = 0.1$, $G/C = 0.02$, $t_p = 15$, and $w = 5$, in units of the NN hopping $t_{\rm nn}$.

\begin{figure}[t]
\includegraphics[width=0.9\columnwidth]{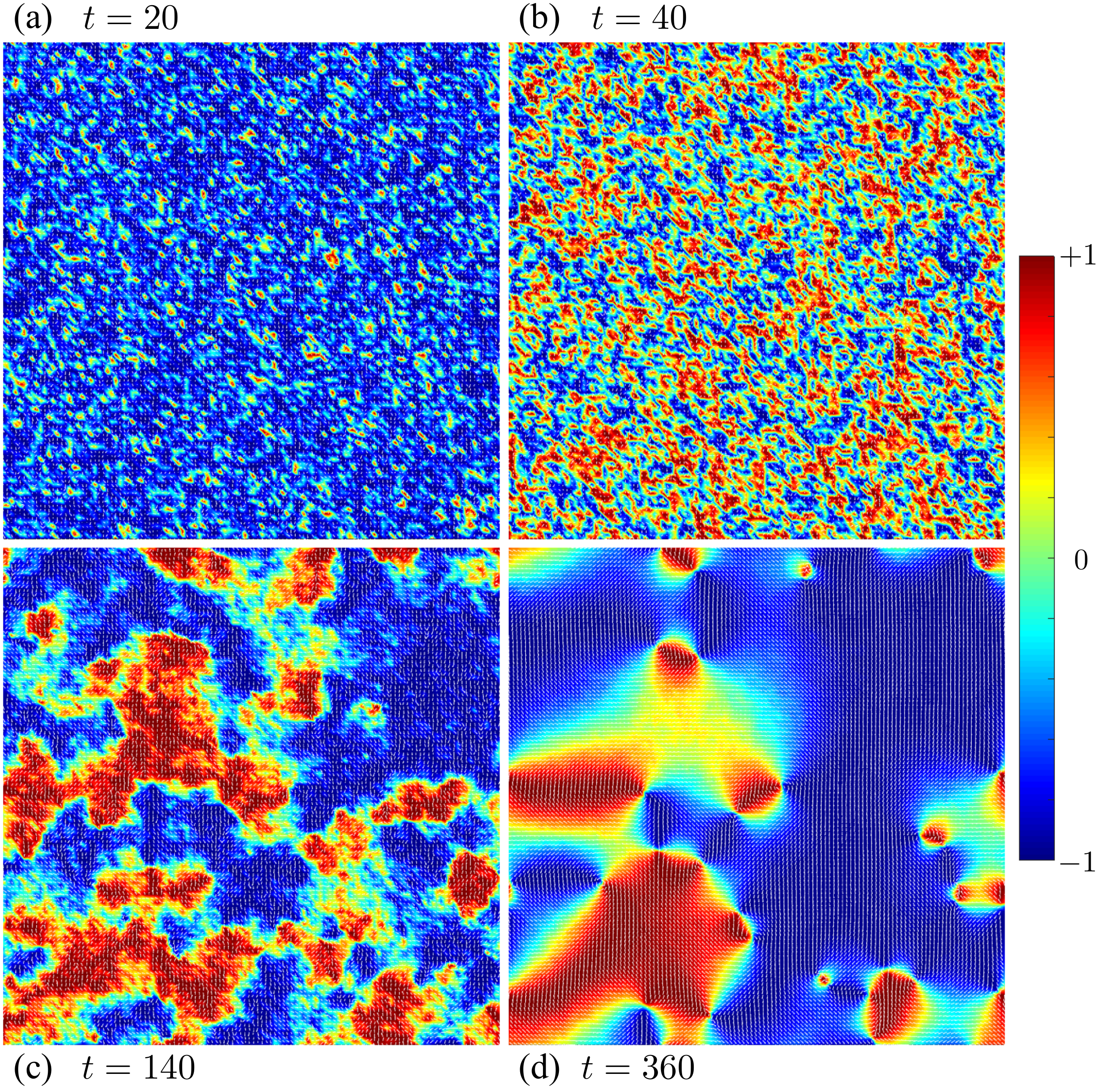}
\caption{(Color online)  
\label{fig:snapshots2} The phase field $\theta(\mathbf r)$ of the SC pairing field $\Delta = |\Delta| e^{i \theta}$ at varying simulation times. The color gradient shows the value of $\sin\theta(\mathbf r)$.
}
\end{figure}

Fig.~\ref{fig:pump-probe}(a) shows the time dependence of the magnitude of the pairing parameter $|\overline \Delta|$ and the electron energy gap $\varepsilon_{\rm gap}$ obtained from simulations. The energy gap is estimated from the instantaneous electron density of states $\rho(\varepsilon)$, shown in Fig.~\ref{fig:pump-probe}(c) for a few representative simulation times. Interestingly, while both quantities exhibit significant drop after the pulse excitation, it seems a very long timescale is required for their recovery even when temperature returns to almost zero. This rather slow dynamics can be attributed to long-lived topological vortex defects of the phase field $\theta(\mathbf r)$, as demonstrated in Fig.~\ref{fig:snapshots2}. Here $\theta(\mathbf r)$ is the phase angle of the SC order parameter, i.e. $\Delta = |\Delta| e^{i \theta}$. The time dependence of the vortex density $n_v$, shown in Fig.~\ref{fig:pump-probe}(b), displays a long tail after the pulse excitation. It is known that phase-ordering of a system when quenched into a symmetry-breaking phase is dominated by defect dynamics~\cite{bray94}. For the $XY$ model, mean-field analysis within the TDGL framework suggests that pair annihilation of defects follows a $t^{-\nu}$ power-law~\cite{toyoki90}, with exponent $\nu = 1$ in 2D. Careful numerical analyses have found logarithmic corrections to this power-law scaling~\cite{yurke93,bray00}. Our preliminary analysis of the SC dynamics suggests a power-law tail with $n_v \sim t^{-0.97}$, which is consistent with known results.

To summarize, we have demonstrated the importance of spatial fluctuations in the nonequilibrium dynamics of superconductivity, especially for pairing field with short coherence lengths. For dissipationless SC dynamics, we have developed a real-space von~Neumann dynamics method within the TDHFB framework. Applying our method to interaction quenches of the negative-$U$ Hubbard model, we have shown that the quench-induced synchronized oscillation of Cooper pairs is unstable against inhomogeneity. Our large-scale simulations seem to confirm the intriguing Cooper pair turbulence scenario that results from the parametric instability of an oscillating pairing field in the large-$U$ regime.
Finally, we have shown that the SC pairing field obeys a Landau-Lifshitz dynamics in the adiabatic limit. By retaining the electron degrees of freedom, large-scale QLLD simulations provide a unique capability to investigate the intriguing interplay of topological defects of the pairing field and the underlying quasiparticles.

\begin{acknowledgements}
The authors thank C. Batista, G. Kotliar, H. Suwa, and Z. Wang for collaboration on related works and insightful discussions. We also thank E. Yuzbashyan, M. Dzero, and B. Altshuler for pointing out the parametric instability of an oscillating pairing field in the large coupling limit. G.-W. C. acknowledges support from the Center for Materials Theory as a part of the Computational Materials Science  (CMS) program, funded by the  DOE Office of Science, Basic Energy Sciences, Materials Sciences and Engineering Division. K. B. acknowledges support from the Laboratory Directed Research and Development (LDRD) program at LANL.
\end{acknowledgements}

\end{document}